\documentclass[12pt]{iopart}
\usepackage{graphicx}
\usepackage{cite}
\usepackage{caption}
\begin{document}

\title[]{A one-dimensional mixing model for the impact of ablative Rayleigh-Taylor instability on compression dynamics}
\author{Dongxue Liu$^1$, Tao Tao$^1$, Jun Li$^{1,3}$, Qing Jia$^1$, Rui Yan$^{2,3}$, Jian Zheng$^{1,3}$}
\address{$^1$Department of Plasma Physics and Fusion Engineering, University of Science and Technology of China, Hefei 230026, People’s Republic of China
}
\address{$^2$Department of Modern Mechanics, University of Science and Technology of China, Hefei 230027, People’s Republic of China}
\address{$^3$Collaborative Innovation Center of IFSA, Shanghai Jiao Tong University, Shanghai, 200240, People’s Republic of China}

\ead{jzheng@ustc.edu.cn}
\vspace{10pt}
\begin{indented}
\item[]February 2025 
\end{indented}

\begin{abstract}
A one-dimensional mixing model, incorporating the effects of laser ablation and initial perturbations, is developed to study the influence of ablative Rayleigh-Taylor instability on compression dynamics. The length of the mixing region is determined with the buoyancy-drag model[arXiv:2411.12392v2 (2024)]. The mixing effect on laser ablation is mainly described with an additional heat source which depends on turbulent kinetic energy and initial perturbation level through a free multiplier.
The model is integrated into a one-dimensional radiation hydrodynamics code and validated against two-dimensional planar simulations. The further application of our model to spherical implosion simulations reveals that the model can give reasonable predictions of implosion degradation due to mixing, such as lowered shell compression, reduced stagnation pressure, and decreased areal density, etc. It is found that the time interval between the convergence of the main shock and stagnation may offer an estimate of mixing level in single-shot experiments.
\end{abstract}

%

\section{Introduction}
Laboratory fusion ignition \cite{abu2022lawson,abu2024achievement} has been successfully achieved at the National Ignition Facility (NIF) through the central ignition scheme\cite{lindl1995development,hurricane2023physics}. In this scheme, 
a hot spot is generated in the center of the fuel shell, a spark of nuclear fusion is initiated, and a burning wave propagates through the shell when the ignition threshold is surpassed. Both the formation of the hot spot and the propagation of the burning wave are closely related to the dynamics of compression\cite{nuckolls1972laser,theobald2014time,tommasini2020time}. Extensive experimental studies have observed compression degradation\cite{robey2013effect,thomas2020deficiencies,yang2024experimental,michel2017measurement,johnson2012neutron,frenje2013diagnosing,landen2020yield,meaney2020diagnostic}, as quantified by key parameters such as adiabat, areal density, and inflight shell thickness. To understand this degradation, researchers have investigated the underlying physical mechanisms in various phases of implosion.

The compression dynamics is significantly influenced by hydrodynamic instabilities across various scales. Low-mode drive asymmetries \cite{scott2013numerical,rinderknecht2020azimuthal}, originating from hohlraum geometry and cross-beam energy transfer (CBET) \cite{kritcher2018energy,edgell2021nonuniform}, have been identified as primary contributors to reduced fuel compression at stagnation. High-mode non-uniformities at fuel-ablator interface \cite{cheng2016effects,amendt2021entropy} enhance fuel entropy through mixing, thereby reducing compression in the acceleration phase, a process exacerbated by preheating effects\cite{li2022mitigation,li2024effect,jones2017progress,robey2012shock}. Moreover, high-mode ablative Rayleigh-Taylor instability (ARTI), seeded by target defects \cite{pak2023overview,divol2024thermonuclear} and laser imprint \cite{goncharov2006early,liu2022mitigating}, fosters the mixing of hot and cold plasmas, thus decreasing compression by affecting laser ablation prior to the onset of these instabilities.

Because of the vast parameter space associated with high-dimensional simulations \cite{clark2016three,clark2019three}, previous studies have employed one-dimensional (1D) simulations that utilize various adjustable multipliers to investigate the effects of mixing in various phases. For example, the fall-line mix model \cite{welser2008application} modifies the fusion process by introducing an adjustable multiplier that calibrates the width of fully atomized mixing region during the deceleration phase. Moreover, at CH-DT interface, plasma flows within the mixing region, as refined by the buoyancy-drag (BD) model\cite{dimonte2000spanwise}, have been modified using either a surface-to-volume (AOV) multiplier \cite{bachmann2022measurement,bachmann2023measuring} or through 1D Reynolds-averaged Navier-Stokes equations\cite{xiao2020modeling,PhysRevE.105.045104} with coefficients tailored for turbulent phases. However, these existing models are not readily applicable to ARTI due to their insufficient treatment of laser ablation effects and initial perturbations.

In this paper, we develop a 1D mixing model that incorporates the effects of laser ablation and initial perturbations to study the impact of ARTI on compression dynamics. The length of the mixing region is calculated using a BD model \cite{liu2024extended} with a time-varying drag coefficient. The mixing effect on laser ablation is characterized through a perturbed heat source, parameterized by turbulent kinetic energy and initial perturbation level via a free multiplier.
The model is then integrated into a one-dimensional radiation hydrodynamics code and validated against two-dimensional planar simulations. Subsequent application to spherical implosion simulations reveals that the model can give reasonable predictions of implosion degradation due to mixing, such as lowered shell compression, reduced stagnation pressure, decreased areal density, and shortened time interval between the convergence of the main shock and stagnation. Notable, this interval may offer an estimate of mixing level in single-shot experiments. 

The paper is organized as follows. We elaborate the 1D mixing model based on simulations in Section 2. We implement the model into the code MULTI-IFE \cite{ramis2016multi} and validate it against FLASH \cite{fryxell2000flash} simulations in planar geometry in Section 3. We apply the model to implosion dynamics analysis in Section 4. Finally, we draw our conclusions in section 5.
\section{1D mixing model based on simulations}
In inertial confinement fusion (ICF), high-mode nonlinear ARTI induces significant mixing between hot and cold plasmas, profoundly affecting laser ablation process. This phenomenon is characterized by an increased characteristic scale length of the ablation front, as evidenced by FLASH \cite{fryxell2000flash} simulations. To investigate this effect in the mixing region refined by the BD model, we develop a 1D mixing model that uses a perturbed heat source to calculate the mixing effects on laser ablation. 
\begin{figure}[h]
\centering
\includegraphics[width=0.7\linewidth]{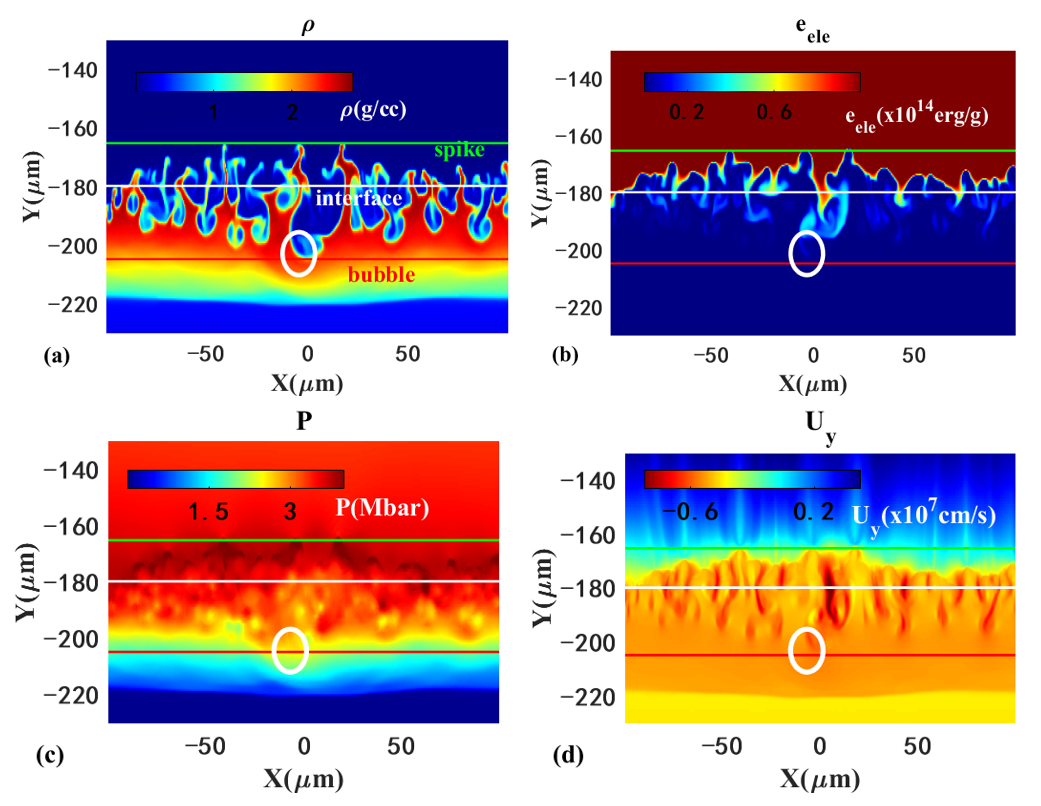}%
\noindent{\caption{\label{fig_1} The distributions of (a) density $\rho$ , (b) specific internal energy of electrons ${{e}_{ele}}$, (c) pressure $P$ and (d) velocity $U_{y}$ , where the white ellipses represent large bubbles. The simulation setup is as follows: The vertically irradiated laser pulse, is square-shaped with a rise time of 0.1 ns and a peak intensity of $25\,\mathrm{TW/cm}^{2}$. The CH planar target has a density of 1 g/cc, a thickness of 90 $\mu \mathrm{m}$, and a length of $L_x=200\,\mu\mathrm{m}$. The multi-mode velocity perturbations, introduced as seeding sources adjacent to the surface, are defined as ${{V}_{p}}(x)=\sum {{V}_{pk}}cos\left(m{{k}_{L}}x+{{\psi }_{k0}} \right)$, ${{k}_{L}}=2\pi /{{L}_{x}}$, where $m$ is an integer ranging from 4 to 10, ${{\psi }_{k0}}$ is a random phase uniformly distributed between zero and one, ${V}_{pk}={V}_{pk0}{{e}^{(-m{{k}_{L}}|y-{{y}_{0}}|)}}$, ${{V}_{pk0}}=C{{\left(m{{k}_{L}} \right)}^{-2}}$, and $C$ is a constant.}}%
 \end{figure} 

We begin by analyzing the influence of ARTI on flow fields. As dipicted in figure \ref{fig_1}, the mixing region is defined as the zone between the bubble front and the spike front. Large bubbles, indicated by white ellipses, demonstrate a notable decrease in density in figure \ref{fig_1}(a), and an obvious increase in the specific internal energy of electrons in figure \ref{fig_1}(b), pressure in figure \ref{fig_1}(c) and velocity in the Y direction in figure \ref{fig_1}(d) compared to ideal uniform distributions in the X direction. These approximate uniform distributions in the X direction motivate our development of a 1D mixing model based on spatially averaged hydrodynamic equations. These equations encompass the mass conservation equation, the momentum equation and the internal energy equation, which are expressed as follows:
\numparts
\begin{eqnarray}
 \frac{\partial \rho }{\partial t}+\frac{\partial \rho {{U}_{j}}}{\partial {{x}_{j}}}=0,\label{1a} \\
 \frac{\partial \rho {{U}_{i}}}{\partial t}+\frac{\partial \rho {{U}_{i}}{{U}_{j}}}{\partial {{x}_{j}}}=-\frac{\partial P}{\partial {{x}_{i}}},\label{1b} \\
\frac{\partial \rho e}{\partial t}+\frac{\partial \rho {U}_{j}e}{\partial {x}_{j}}=-P\frac{\partial {U}_{j}}{\partial {{x}_{j}}}-\frac{\partial q_j}{\partial {{x}_{j}}},\label{1c} 
\end{eqnarray}
\endnumparts
where $P$ represents pressure, $e$ denotes specific internal energy, and $\vec{q}=\vec{q}_{ele}+\vec{q}_{rad}$ is the total heat flux that has a radiation and electron conductivity component. The electron conductivity is given by ${q}_{ele,j}=-\kappa_{ele}\frac{\partial T_{ele}}{\partial {{x}_{j}}} ,$ where $\kappa_{ele}$ is the coefficient of electron conductivity, proportional to $T_{ele}^{5/2},$ and $T_{ele}$ represents the electron temperature.

The averaging process involves a spatial average ($ \rho =\bar{\rho} +\rho ' $) and a mass-weighted average ($ U=\overline{\overline{U}}+U'' $). The relationship between these two averages is characterized by \[ \overline{\overline{U}}=\frac{\overline{\rho U}}{\bar{\rho} }=\overline{ U} +\frac{\overline{\rho 'U'}}{\bar{\rho} } , \bar{\rho} =\int\limits_{0}^{L_{x}}{\rho dX}/\int\limits_{0}^{L_{x}}{dX} ,\] where $ \bar{\rho} $ and $ \rho ' $ represent the mean and fluctuating density of the spatial average, while $ \overline{\overline{U}} $ and $ U'' $ denote the mean and fluctuating velocity of the mass-weighted spatial average. Upon averaging, equation (\ref{1a}) transforms into the form as one-dimensional mass conservation equation: \begin{equation}
\frac{\partial \bar{\rho} }{\partial t}+\frac{\partial \bar{\rho} {{{\overline{\overline{U}}}}_{y}}}{\partial {y}}=0.\label{2} 
\end{equation}Similarly, equation (\ref{1b}) becomes 
\begin{equation}
 \frac{\partial \bar{\rho} {{{\overline{\overline{U}}}}_{y}}}{\partial t}+\frac{\partial \bar{\rho} {{{\overline{\overline{U}}}}_{y}}{{{\overline{\overline{U}}}}_{y}}}{\partial {y}}=-\frac{\partial \overline{ P}}{\partial {y}}-\frac{\partial \overline{\rho {{U}_{y}}''{{U}_{y}}''} }{\partial {y}}. \label{3} 
\end{equation}
The right-hand side of equation (\ref{3}) consists of two components: the pressure gradient and the shear force associated with the velocity gradient. The shear force serves as the source of turbulent kinetic energy, defined as $ {K}_{f}=\frac{1}{2}\frac{\overline{\rho U_i^{''}U_i^{''}}}{\overline{\rho}} $. During the implosion process, the inflight shell moves on the order of hundreds of micrometers, while the bubble's movement is limited to few tens of micrometers to maintain shell integrity. Consequently, the perturbed velocity constitutes at most 10\% of the implosion velocity, and $ {K}_{f}$ accounts for less than 1\% of the total kinetic energy $\overline{\overline{E}}_{k}$. This context allows us to neglect the influence of the shear force , which is further supported by the larger pressure gradient than the velocity gradient in the Y direction near the ablation front, as illustrated in figure \ref{fig_1}(c) and (d).

Moreover, disregarding the correlations in $\vec{q}_{rad}$, the averaged internal energy equation can be expressed as follows:
\begin{equation}
\frac{\partial \bar{\rho}\bar{\bar{e}} }{\partial t}+\frac{\partial \bar{\rho} \overline{\overline{U}}_{y}\bar{\bar{e}}}{\partial y}=-\overline{P}\frac{\partial \overline{\overline{U}}_{y}}{\partial {y}}-\frac{\partial \bar{q}_{rad,y}}{\partial {y}}-\frac{\partial{\kappa_{ele}(\overline{\overline{T}}_{ele})\frac{\partial \overline{\overline{T}}_{ele}}{\partial {y}}}}{\partial {y}}+W,\label{4}
\end{equation}
where $W$ is expressed as \[W=-\frac{\partial \overline{\rho{U}_{y}^{''}e^{''}}}{\partial y}-\overline{P\frac{\partial {U}_{y}^{''}}{\partial {y}}}-(\frac{\partial{\bar{q}_{ele,y}}}{\partial{y}}-\frac{\partial{\kappa_{ele}(\overline{\overline{T}}_{ele})\frac{\partial \overline{\overline{T}}_{ele}}{\partial {y}}}}{\partial {y}}),\] including correlations from diffusion, compression work and thermal conduction. The term in bracket with the highest degree of $T_e^{7/2}$ presents substantial challenges in developing accurate closure models. Therefore, recognizing that equation (\ref{2}) and (\ref{3}) share the same form as one-dimensional equations when neglecting the shear force, we develop a phenomenological mixing model to address $W$ in equation (\ref{4}). 

\begin{figure}[h]
\centering
\includegraphics[width=0.6\linewidth]{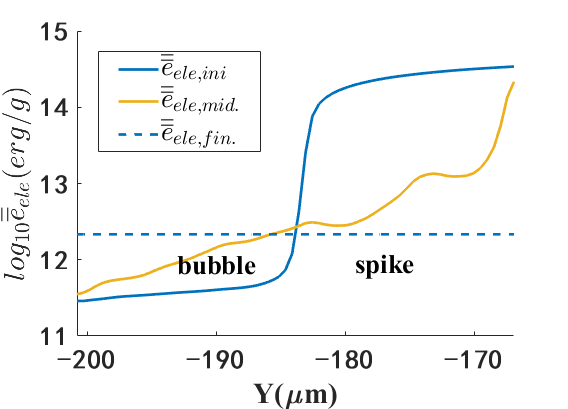} \caption{\label{fig_2} The distributions of ${\bar{\bar e}_{ele}}$ at the same time,including the initial state ${\bar{\bar e}_{ele,ini.}}$, the middle mixing state calculated from figure \ref{fig_1}(b), and the final state of isothermal mixing ${\bar{\bar e}_{ele,fin.}}=\int{\bar{\rho} {\bar{\bar e}_{e,ini.}}dV}/\int{\bar{\rho} dV}$, where $V$ represents the volume of the target. The differences among these three states arise from various mixing levels.}%
 \end{figure}
The spatial extent of the mixing model is determined based on the average characteristic of $ {{e}_{ele}} $, as represented by the yellow line in figure \ref{fig_2}. This average characteristic indicates that mixing enhances the characteristic scale length of the ablation front, leading to an evolution from the initial state $ {{\bar{\bar{e}}}_{ele,ini.}} $ to the middle mixing state $ {{\bar{\bar{e}}}_{ele,mid.}} .$ Notably, $ {{\bar{\bar{e}}}_{ele,mid.}} $ at both the spike front and the bubble front closely resembles $ {{\bar{\bar{e}}}_{ele,ini.}} $ , underscoring the mixing region as the primary area influenced by ARTI. Therefore, the mixing region is calculated using the BD model\cite{liu2024extended}, expressed as follows:
\begin{equation}
\frac{d{{V}_{B}}}{dt}={{A}_{t}}g-D\frac{{{V}_{B}}\left| {{V}_{B}} \right|}{{{h}_{B}}},\label{5}
\end{equation}
 where $D(t),\,A_t(t)$ and $g(t)$ represent the time-varying drag coefficient, Atwood number and acceleration, respectively. The center of the mixing region is located at the ablation front, extending a length of $h_B$ above and below. 
 
The computation of $W$ in equation (\ref{4}) is based on the relaxation of a perturbed heat source. We assume that the system evolves toward a final isothermal state, denoted as \( \bar{\bar{e}}_{ele,fin.}(t, \Delta t). \) In the absence of ablative flux, characterized by $V_a=0$ and $K_f/\overline{\overline{E}}_k=1,$ the isothermal state can be achieved over a relaxation time $\Delta t$, where $V_a$ represents the averaged ablative velocity. The maximum perturbed heat source is defined as \[\Delta {\bar{\bar e}_{ele}(t,\Delta t,y)}={\bar{\bar e}_{ele,fin.}(t,\Delta t)}-{\bar{\bar e}_{ele,ini.}(t,y)},\]which exhibits positive values near bubbles and negative values near spikes, as illustrated in figure \ref{fig_2}. Consequently, the perturbed heat source in equation (\ref{5a}) is formulated as $\aleph(t,y) \Delta {{\bar{\bar{e}}}_{ele}(t,\Delta t,y)}$ and is utilized to close $W$: 
\numparts
\begin{eqnarray}
W=\frac{\partial Q_{per.}(t,y)}{\partial t}=\frac{\aleph(t,y) \Delta {{\bar{\bar{e}}}_{ele}(t,\Delta t,y)}}{\lambda/V_{a,loc}(t)},\label{5a}\\
\aleph(t,y)=\sqrt{K_f(t,y)/\overline{\overline E}_{k}(t,y)}\times{\xi(t)}^{-1/2(-1+sign(\Delta {{{\bar{\bar{e}}}}_{ele}(t,y)}))}, 
\label{5b}\\
{{K}_{f}(t,y)}={\frac{1}{2}{f}_{mix}}V_B(t)^2(1-{{y}^{2}}/h_{B}^{2}),\label{5c}\\
 {\int\limits_{V\in mix\,region}W\times\rho dV }=0.\label{5d}
\end{eqnarray}
\endnumparts
Here, the relaxation time of the heat source is defined as $\lambda/V_{a,loc}(t),$ where $\lambda$ is the dominant perturbation wavelength at the onset of acceleration and the local ablative velocity $V_{a,loc}(t)$ approximates the ablative velocity from 1D simulations. The amplitude of the heat source is associated with the mixing level, represented by $\aleph(t,y)$ in equation (\ref{5b}), where the $sign$ function is defined as: \[sign(x<0)=-1,sign(x>0)=1,sign(x=0)=0.\] Additionally, a specific spatial distribution of $ {{K}_{f}} $ is assumed, similar to that used in the RANS model \cite{xiao2020modeling}, as expressed in equation (\ref{5c}).  $\xi$(t) in equation (\ref{5d}) ensures that the integral of the perturbed heat source within the mixing region equals zero. 

In equation (\ref{5b}), $\aleph(t,y)$ is  proportional to the percentage of specific turbulent kinetic energy, expressed as $\sqrt{K_f(t,y)/\overline{\overline{E}}_k(t,y)},$ whose function form may vary. Although $K_f(t,y)$ does not evolve according to equation (\ref{3}) in the absence of the shear force, its maximum value can be defined as $\frac{1}{2} f_{mix} V_B^2(t)$, occurring near the interface between bubbles and spikes, specially at the ablation front, as illustrated in figure \ref{fig_1}(d). Here, $ {{f}_{mix}} $ is an adjustable multiplier that characterizes the initial perturbation level at the onset of acceleration. For example, if only a local defect is present at the ablation front, $ {{f}_{mix}} $ approaches the lower limit of zero, whereas a fully perturbed ablation front results in $ {{f}_{mix}} $ approaching the upper limit of one. We also recognize that the mixing level can be manipulated in two distinct ways: by changing the time evolution of $h_B$ and by adjusting ${{f}_{mix}}$. The effects of these two ways are obviously different, suggesting the complexity of mixing dynamics at the ablation front. 

Eventually, to address $W$ in equation (\ref{4}), the phenomenological mixing model explicitly calculates the relaxation of the perturbed heat source, as expressed in equation (\ref{7}): 
\begin{equation}
 {\bar{\bar e}_{ele,mid.}(t,dt,y)-\bar{\bar e}_{ele,ini.}(t,y)}=W\times dt,\label{7}
\end{equation}during the lookup process of equation of state (EOS), where $dt$ is the time step. Furthermore, within the mixing region, thermal relaxation promotes the equilibration of temperatures between electrons and ions, resulting in a corresponding 1D mixing model for the mass-weighted spatially averaged specific internal energy of ions. The integration of the model will be detailed in the following section.
\section{Verification
 of the model}
In MULTI-IFE, the hydrodynamic evolution is consistent with equation (\ref{2}-\ref{4}), while incorporating the implementation of equation (\ref{5}), and equation (\ref{7}) is explicitly calculated to describe the correlations in (\ref{4}) prior to the EOS table lookup\cite{ramis2016multi}. When the time evolution of the mixed region is confirmed, we determine $ {{f}_{mix}} $ via ensuring the calculation results consistent with that of 2D simulations.
\begin{figure}[h]
\centering
 \includegraphics[width=0.4\linewidth]{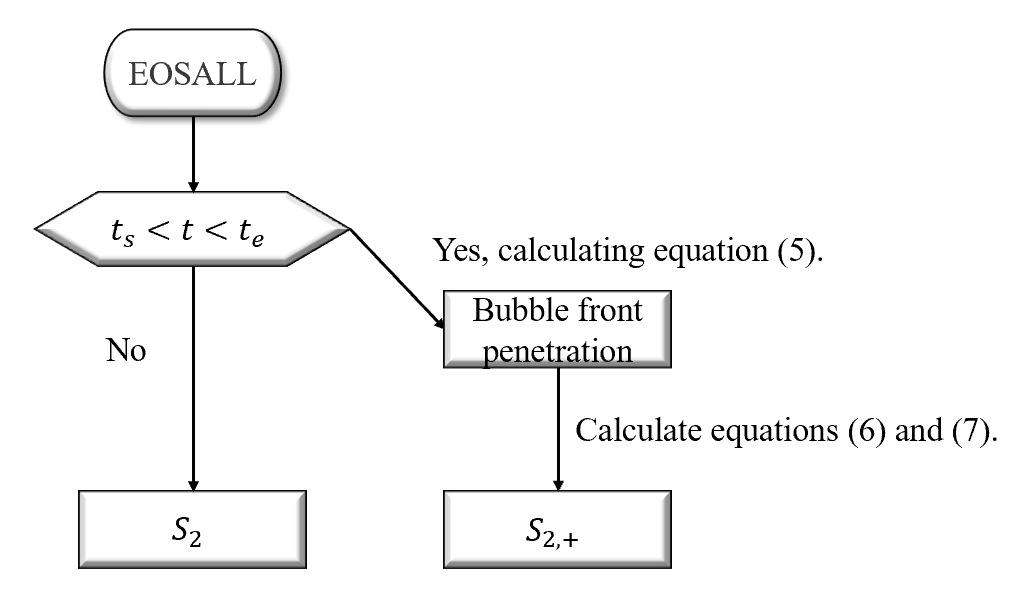}%
 \caption{\label{fig_3} The flow chart of implementation. Here, ${{S}_{2,+}}$ represents the modified $\bar{\bar{e}}$ using equation (6).}%
 \end{figure}

The implementation process begins by defining several global input parameters, such as the initial perturbations, the onset time $ {{t}_{s}} $ when perturbations enter the nonlinear phase\cite{birkhoff1955taylor}, and the end time $ {{t}_{e}} $ marking the end of the acceleration phase. These parameters are essential for calculating $ {{V}_{B}} $ and $ {{h}_{B}} $ using equation (\ref{5}). As illustrated in figure \ref{fig_3}, from $t={{t}_{s}}$ to $t={{t}_{e}}$, $ \bar{\bar{e}} $ is updated according to equation (6) and (\ref{7}), leading to corresponding updates in temperature, pressure and other relevant parameters. By inferring $ {{f}_{mix}} $, these updated parameters enhance the consistency of the mixing model with the averaged results obtained from two-dimensional simulations.

\begin{figure}[h]
 \centering
 \includegraphics[width=1.0\linewidth]{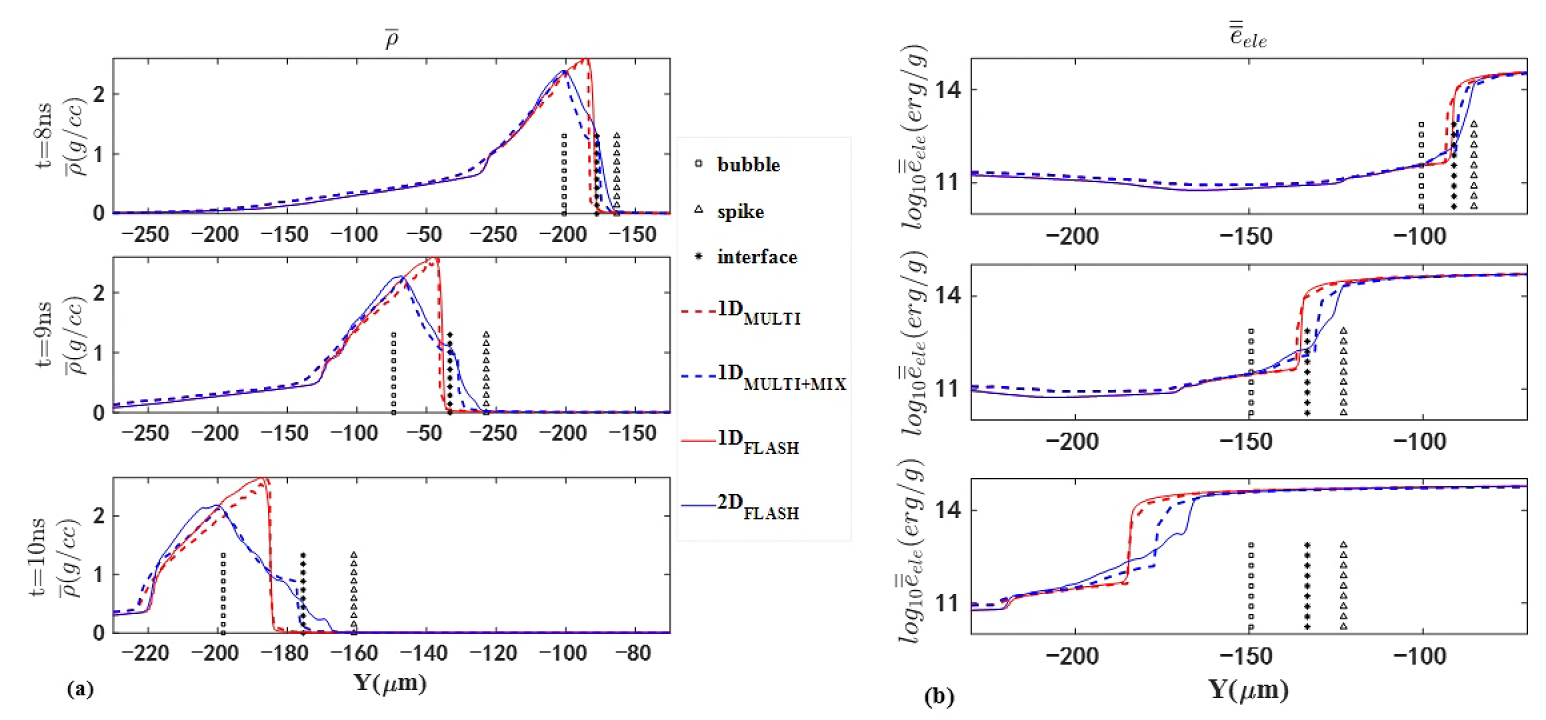}%
 \caption{\label{fig_4}The distributions of $\overline{\rho } $ and ${{\bar{\bar{e}}}_{ele}}$ for cases with (blue) and without (red) velocity perturbations. The solid line is from FLASH, the dashed line is calculated using MULTI-IFE, and the symbol lines from left to right represent the bubble front, the interface between the bubble front and the spike front and the spike front.}%
 \end{figure}

The results of the mixing model applied to laser-ablated planar targets are presented in figure \ref{fig_4}(a) and (b). In these figures, solid lines represent FLASH simulations, while dashed lines correspond to MULTI-IFE simulations. Blue lines denote simulations with initial velocity perturbations, whereas red lines indicate simulations without perturbations. The overlapping red lines imply that the time evolution of \( \bar{\rho} \) and \( \bar{\bar{e}}_{{ele}} \) is similar in both simulations, despite differences in computational cells. In the presence of perturbations, the 2D simulation observes an increase in $ {{\bar{\bar{e}}}_{ele}} $ and a decrease in $ \bar{\rho} $ near the bubble front, while an opposite trend is observed near the spike front. Notably, mixing causes the maximum specific internal energy gradient of electrons to shift outward into the coronal region. Our 1D mixing model effectively captures the time evolution of these characteristics with $ {{f}_{mix}}=0.03, $ however, a notable deviation is observed near the spike front, which is anticipated due to the increased values of $\Delta \bar{\bar{e}}_{ele}$ in this region, as illustrated in figure \ref{fig_2}.

We further quantify the compression of the implosion shell, defined as the region between the internal and external points where the density equals $1/e$ of the maximum density. At $t=10$ ns, the compression of the inflight shell, represented by the average density in planar geometry and mass-weighted adiabat, decreases by approximately $ 50\% $ when mixing is considered. Specifically, the average density and mass-weighted adiabat of the implosion shell from 2D FLASH simulations are 1.58 g/cc and 1.77, while the values calculated by 1D MULTI-IFE are 1.49 g/cc and 1.62. The relative errors for average density (5.7\%) and mass-weighted adiabat (8.5\%) both remain below $10\%$ . 
 \begin{figure}[h]
 \centering
\includegraphics[width=0.3\linewidth]{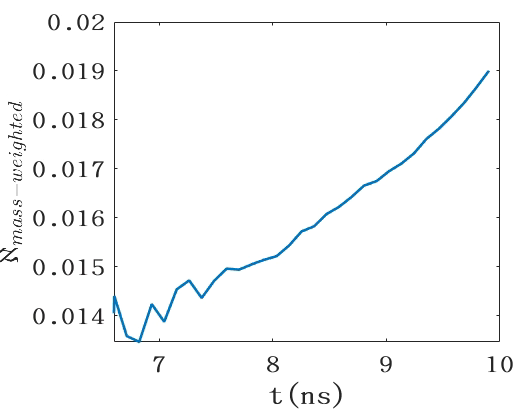}%
 \caption{\label{fig_9} The time evolution of the mass-weighted mixing level $\aleph_{mass-weighted}$. The initial irregular change of mixing level is attributed to deviations in the actual initial perturbations. }%
 \end{figure}
 
Furthermore, We define the mass-weighted mixing level as \[\aleph_{mass-weighted}(t)=\frac{\int\limits_{V\in mix\,region}{\left(\aleph (t,y)\rho dV \right)}}{\int\limits_{V\in mix\,region}{\left(\rho dV \right)}} , \] with its time evolution illustrated in figure \ref{fig_9}. During the development of instabilities, the mixing level, as indicted using $K_f$, increases over time. Ultimately, the relationship between the evolving instabilities and the rising mixing level enhances our comprehension of mixing. 
\section{Application of the model to implosion dynamics
}
 \begin{figure}[h]
 \centering
\includegraphics[width=0.5\linewidth]{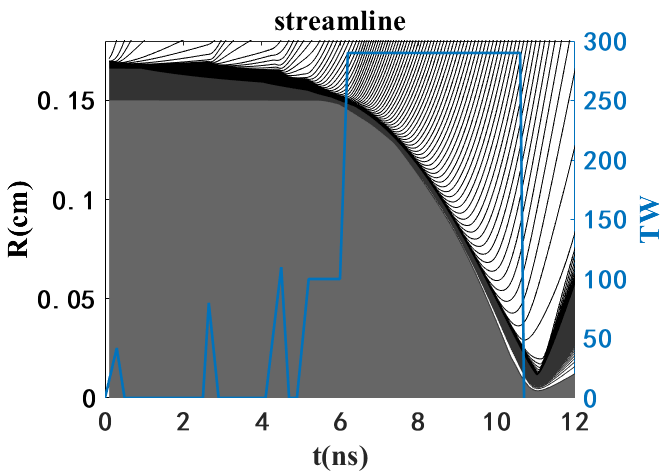}%
 \caption{\label{fig_5} The pulse shape and streamline diagram for a 1.5-MJ triple-picket design.}%
 \end{figure}
Implosion experiments have measured several typical phenomena, including the reduced compression of the inflight shell, lowered stagnation pressure, decreased areal density of the hot spot and early onset of bang time \cite{rygg2007time,robey2016performance}. To investigate the relationship between these phenomena and mixing, we implement the model in spherical geometry. In this context, $ \bar{\bar{e}}_{ele,fin.}$ is calculated using the formula:\[\bar{\bar{e}}_{ele,fin.}=\int \limits_{R}4\pi {R}^2\bar{\rho} \bar{\bar{e}}_{ele,ini.}dR/\int\limits_{R}4\pi {R}^{2}\bar{\rho}dR.\] Figure \ref{fig_5} presents the pulse shape and streamline diagram for a 1.5-MJ triple-picket design \cite{craxton2015direct}without the mixing model. The typical target scheme is reflected through grayscale: the outer plastic has a density of 1 g/cc, the middle DT ice has a density of 0.25 g/cc, and the inner DT gas has a density of 6 mg/cc. The inflight shell begins to accelerate at $t=5.5$ ns, continuing until the main shock reflects from the center to the interior of the inflight shell at $t=10.5$ ns, marking the onset of the deceleration phase. At approximately 11 ns, the hot spot stagnates, characterized by its minimum volume.
\begin{figure}[h]
\centering
\includegraphics[width=0.6\linewidth]{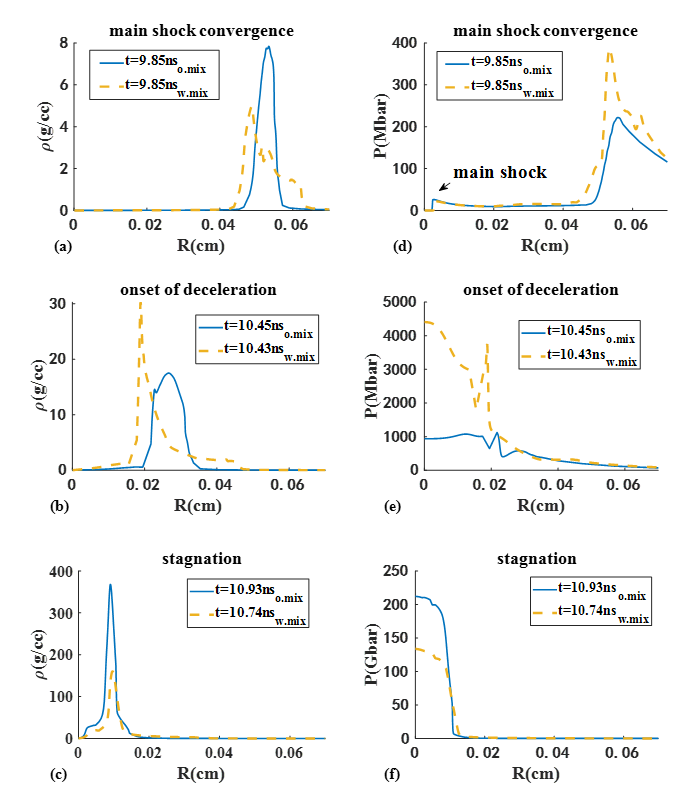}%
 \caption{\label{fig_7} The spatial distributions of density and pressure at three moments: the convergence of the main shock in (a) and (d), the onset of deceleration phase in (b) and (e), and the stagnation of the hot spot in (c) and (f). The hot spot density and stagnation pressure are both reduced due to mixing. }%
 \end{figure}
 
During the acceleration phase from 5.5 ns to 10.5 ns, we employ the mixing model at the ablation front with $ f_{mix}=0.03 $ , as calibrated in Section 3. 
The quality of the hot spot at three characteristic moments are detailed in figure \ref{fig_7}. The outer radius of the hot spot is denoted by the maximum density gradient. In figure \ref{fig_7}(a), the mixing model predicts a decreased inner radius\textsuperscript{[11]} accompanied by an increased thickness of the inflight shell, indicative of decreased compression. Notably, the convergent time of the main shock, shown in figure \ref{fig_7}(d), remains unaffected by mixing, because it occurs with only minor perturbations at the onset of acceleration. The combination of reduced hot spot radius and spherical convergence geometry results in enhanced density and pressure within the hot spot at the onset of deceleration, as demonstrated in figures \ref{fig_7}(b) and (e). Differently, the expanded radius caused by mixing effects leads to decreased density and pressure within the hot spot at stagnation, as shown in figures \ref{fig_7}(c) and (f). Moreover, Changes in implosion dynamics caused by mixing contribute to an earlier onset of the deceleration phase and the stagnation. These phenomena suggest that mixing may lead to an earlier bang time\textsuperscript{[42}\textsuperscript{, }\textsuperscript{43]}.
 \begin{figure}[h]
 \centering
 \includegraphics[width=1.0\linewidth]{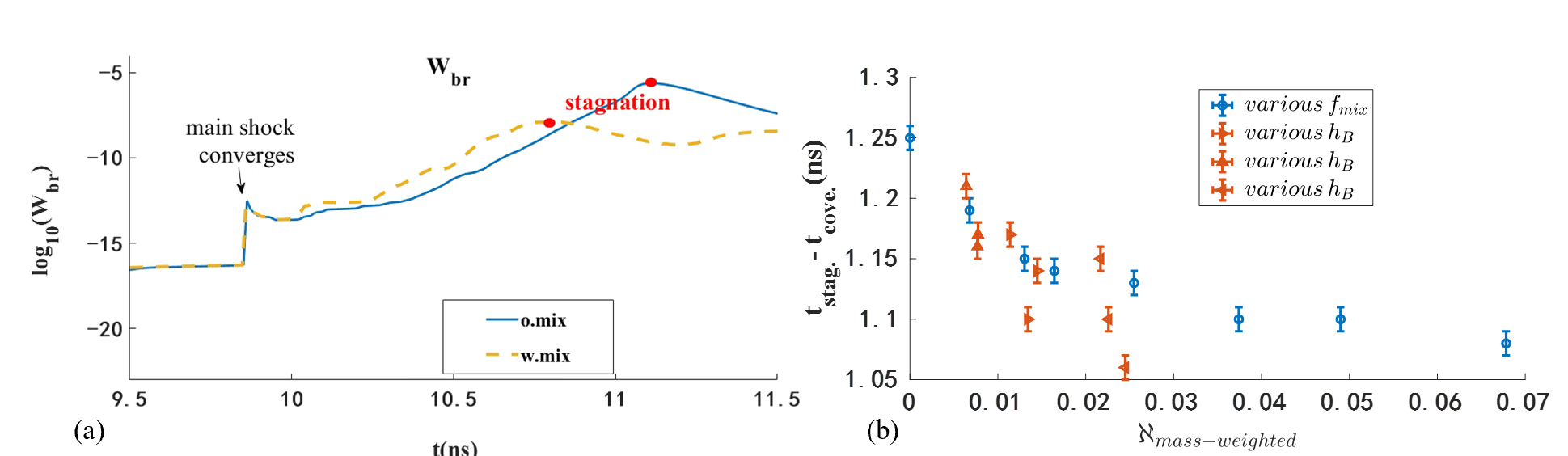}%
 \caption[width=1.0\linewidth]{(a)The total bremsstrahlung energy of the hot spot within a radius of 10 $\mu\mathrm{m}$ . The amplitude of self-emissivity and the interval between the convergence of the main shock and stagnation, expressed as $t_{stag.}-t_{cove}$, are reduced under the influence of mixing. (b) The relationship between $\aleph_{mass-weighted} $ at the end of acceleration and $t_{stag.}-t_{cove}.$ Blue circles indicate simulations where $\aleph_{mass-weighted} $ is varied using ${f}_{mix}$, while orange triangles represent changes in $\aleph_{mass-weighted} $ achieved through different time evolutions of $ {h}_{B}$. Triangles of the same orientation correspond to the same $f_{mix}$. The error bars reflect the uncertainty associated with the time steps in the simulations.}
 \label{fig_8}%
 \end{figure}

The self-emissivity signal of the hot spot can indicate the moment of stagnation and the quality of the hot spot, making it an indicator of the mixing level. We present the total bremsstrahlung energy of the hot spot within a radius of 10 um, i.e., \[ {{W}_{br}}\sim \int{{{\rho }^{2}}{{T}_{e}}^{1/2}dV}.\] As seen in figure \ref{fig_8} (a), the peak of X-ray self-emission with mixing is much lower than that without mixing. Furthermore, 
the interval between the convergence of the main shock and stagnation is reduced. We then conducted simulations to investigate the relationship between the interval and the mass-weighted mixing level at the end of acceleration, denoted as $\aleph_{mass\,weighted}.$ As illustrated in figure \ref{fig_8} (b), when $\aleph_{mass\,weighted}$ is varied using two different methods, both approaches reveal that the interval decreases as the mixing level increases, albeit at different extent. This decreasing trend may offer an estimate of mixing level in single-shot experiments.
\section{Conclusions}
To investigate the influence of ARTI on compression dynamics within a reduced parameter space, we present a one-dimensional mixing model that incorporates the effect of laser ablation and initial perturbations. The length of the mixing region is determined with a BD model. The mixing effect on laser ablation is mainly described with an additional heat source which depends on turbulent kinetic energy and initial perturbation level through a free multiplier. Following the successful implementation of the model, we calibrate the adjustable multiplier against two-dimensional planar simulations. Furthermore, the application of our model to spherical implosion simulations reveals that the model can give reasonable predictions of implosion degradation due to mixing, such as lowered shell compression, reduced stagnation pressure, and decreased areal density, etc. It is found that the time interval between the convergence of the main shock and stagnation may offer an estimate of mixing level in single-shot experiments. Overall, our model could serve as a valuable tool for evaluating the impact of ARTI on implosion dynamics, particularly for future schemes aimed at enhancing compression, such as the SQ-n design.\cite{tommasini2023increased,clark2022exploring}. 

\section*{acknowledgments}
This work is supported by the supported by the National Key R and D Projects (Grant No. 2023YFA1608400), Strategic Priority Research Program of the Chinese Academy of Sciences (Grant No. XDA25010200), and Nature Science Foundation of China (Grant No. 12375242).
\section*{Data Availability Statement}
All data that support the findings of this study are included within the article (and any supplementary files).
\section*{References}
\bibliography{iopart-num}

\end{document}